\documentclass{elsart}

\usepackage{graphicx}

\usepackage{amssymb}

\def\ds{\displaystyle}	
\makeatletter
 
  \@addtoreset{equation}{section}
 \makeatother

\begin{document}

\begin{frontmatter}



\title{Entropy of capacities on lattices and set systems}


\author[label1,label2]{Aoi Honda\corauthref{cor1}}
\ead{aoi@ces.kyutech.ac.jp}
\corauth[cor1]{corresponding author}
\author[label2] {Michel Grabisch} 
\address[label1]{Kyushu Institute of Technology}
\address[label2]{University Paris I-Panth\'{e}on-Sorbonne}

\begin{abstract}
We propose a definition for the entropy of capacities defined on lattices.
Classical capacities are monotone set functions and can be seen as a
generalization of probability measures. Capacities on lattices address the
general case where the family of subsets is not necessarily the Boolean lattice
of all subsets. Our definition encompasses the classical definition of Shannon
for probability measures, as well as the entropy of Marichal defined for
classical capacities. Some properties and examples are given.
\end{abstract}

\begin{keyword}
entropy \sep capacity \sep lattice \sep regular set system \sep convex geometry \sep antimatroid
\end{keyword}
\end{frontmatter}

\section{Introduction}
The classical definition of Shannon for probability measures is at the core of
information theory. Therefore, many attempts for defining an entropy for set
functions more general than classical probability measures have been done, in
particular for the so-called \emph{capacities} \cite{cho53} or \emph{fuzzy
measures} \cite{sug74}. Roughly speaking, capacities are probability measures
where the axiom of additivity has been replaced by a weaker one, monotonicity
with respect to inclusion. 

First definitions of an entropy for a capacity were proposed independently 
and approximately at the same time by Yager \cite{Yager:1999,Yager} and 
Marichal and Roubens \cite{M:2002,MR:1999,MR}. 
The idea of Yager was to compute the Shannon entropy of 
the Shapley value of a capacity.  To make the
discussion more precise, let us 
consider a finite universal set $N$, and a
capacity $v$ defined on it. The Shapley 
value \cite{sha53} is a notion coming
from cooperative game theory, and can be seen 
as a probability distribution
$\phi$ over $N$ which represents the average contribution 
of each element $i\in
N$ in the value of $v$, that is, $v(S\cup i) - v(S)$, for all subsets 
$S$ of $N\setminus i$. 
A slightly different proposition was done by Marichal and Roubens \cite{MR}, just by
changing the place of the function $h(x):=-x\log x$. It turned out that this
definition seemed to be the right one, with properties close to the classical
Shannon entropy \cite{Shannon}. 
In particular, it is strictly increasing towards 
the capacity which maximizes entropy. 
An important result, due to Dukhovny \cite{Dukhovny}, and also independently found 
by Kojadinovic et al. \cite{KM},showed that the definition of Marichal
and Roubens could be written as the average of classical entropy along maximal
chains of the Boolean lattice of subsets of $N$. 


In this paper, we consider yet more general functions than capacities, in the
sense that the underlying system of sets may be not the whole collection of
subsets of $N$, but only a part of it, provided that this collection forms a
lattice. This is motivated partly by cooperative game theory, where $N$ is the
set of players, subsets are called coalitions, and the fact that all subsets may
not belong to the set systems corresponds to the situation where some coalitions
may be forbidden. This is considered for example by Faigle and Kern \cite{FK} (games
with precedence constraints). Our approach will follow Dukhovny, in the sense
that our basic material will be the maximal chains over the considered lattice,
and we will try to make the least possible assumptions on the lattice in order
that our construction works. This permits to consider our definition in a more
abstract way, forgetting about the corresponding set system, and working only on
the lattice. In this way, it is possible to consider as particular cases
bi-cooperative games of Bilbao \cite[Section 1.6]{bifejile00}, and multichoice games
\cite{hsra93}. 

Section 2 recalls classical facts on Shannon's entropy and the definition of
Marichal and Roubens, Section 3 gives the necessary material for lattices and
convex geometries, while Section 4 introduces the notion of capacity on a
lattice, viewed as a set system. Section 5 gives the definition of entropy for
such capacities on lattices, and studies its properties. Section 6 gives
examples of different lattices, so as to recover well known cases.

\section{Entropy of classical capacity}
Throughout this paper, we consider a finite universal set $N=\{1, 2, \ldots,
n\}$, and $2^N$ denotes the power set of $N$. Let us consider $\mathfrak{S}$ a
subcollection of $2^N$. 
Then we call $(N,\mathfrak{S})$ (or simply $\mathfrak{S}$
if no ambiguity occurs) a \emph{set system}. 
In the following, $(N,\mathfrak{S})$ or simply $\mathfrak{S}$ 
will always denote a set system. 

\begin{defn}[capacity] Let $(N, \mathfrak{S})$ be a set system, 
with $\emptyset, N \in \mathfrak{S}$. 
{\rm (i)}\ A {\it game} is a set function $v: \mathfrak{S} \to \mathbb{R}$ which satisfies 
$v(\emptyset)=0$. 
{\rm(ii)}\ A set function $v : \mathfrak{S} \to
[0, 1]$ is a capacity if it satisfies  that $v(\emptyset)=0$, $v(N)=1$, 
and $v(A)\leqq v(B)$ whenever $A\subseteq B$. 
\end{defn}

Usually classical games and capacities are defined on $(N, 2^N)$. 

\begin{defn}[Shapley value] 

The Shapley value of a capacity $v$ is defined by
$$\phi(v):=(\phi_1(v), \ldots, \phi_n(v))\in [0, 1]^n$$
and 
\begin{equation}\hspace{3cm}\label{sv}
\phi_i(v):=\sum_{A\subseteq N\setminus \{i\}}\gamma_{|A|}^n 
[v(A\cup \{i\})-v(A)], 
\end{equation}
where 
\begin{equation}\label{gamma}
\hspace{4cm}\ds\gamma_{k}^{n}:=\frac{(n-k-1)!k!}{n!}.
\end{equation}
\end{defn}
Remark that $\sum_{i=1}^n \phi_i(v)=1$ holds. 

\begin{defn}[Shannon Entropy\cite{Shannon}] 
Let $p, q$ be probability measures on 
$(N, 2^N)$. The Shannon entropy of $p$ and the relative entropy 
of $p$ to $q$ are defined by 
\begin{eqnarray*}
\hspace{4cm}H_S(p)&:=&\sum_{i=1}^n h[p_i],\\
\hspace{4cm}H_S(p; q)&:=&\sum_{i=1}^n h[p_i; q_i],
\end{eqnarray*}
where $p_i:=p(\{i\}), q_i:=q(\{i\}), \ds h(x):=-x\log x$, and $h(x; y):= x \log\frac{x}{y}$. 
\end{defn}
Here $\log$ denote the base $2$ logarithm and 
by convention $\log 0:=0$. 
\begin{defn}[Marichal's entropy\cite{MR}] 
Let $v$ be a capacity on $(N, 2^N)$. Marichal's entropy of a capacity $v$ is defined by
\begin{equation}\hspace{2cm}\label{me}
H_M(v):=\sum_{i=1}^n \sum_{A\subseteq N\setminus \{i\}}\gamma^n_{|A|} 
h[v(A\cup \{i\})-v(A)],
\end{equation}
where $\ds\gamma_{k}^{n}$ is defined by {\rm (\ref{gamma})}. 
\end{defn}
Remark that equations (\ref{sv}) and (\ref{me}) are similar.  
Dukhovny gives a representation of Marichal's entropy using maximal 
chains of $2^N$ \cite{Dukhovny}. 


\begin{defn}[maximal chain of set system]
Let $(N, \mathfrak{S})$ be a set system, with  
$\emptyset, N\in \mathfrak{S}$. 
If $C=(c_0, c_1, \ldots, c_m)$ satisfies that $
\emptyset=c_0 \subsetneq c_1 \subsetneq \cdots 
\subsetneq c_m =N, c_i \in \mathfrak{S}$ and there is no element 
$c\in \mathfrak{S}$ such that $c_{i-1} \subsetneq c \subsetneq c_{i}$ for any $i\in \{1, \ldots, m\}$ then we call $C$ a maximal chain of 
$\mathfrak{S}$. 
\end{defn}
We denote the set of all maximal chains of $\mathfrak{S}$ by 
$\mathcal{C}(\mathfrak{S})$. 
Let $v$ be a capacity on $\mathfrak{S}$. Define $p^{v, C}$ by 
\begin{eqnarray}
p^{v, C}&:=&(p^{v, C}_1, p^{v, C}_2, \ldots, p^{v, C}_m)\nonumber\\
&=&(v(c_1)-v(c_0), v(c_2)-v(c_1), \ldots, v(c_m)-v(c_{m-1})), \label{dp}
\end{eqnarray}
where $C=(c_0, c_1, \ldots, c_m)\in \mathcal{C}(\mathfrak{S})$. Note that
$p^{v, C}$ is a probability distribution, i.e. $p^{v, C}_i\geqq 0, i=1, \ldots, m$ and $\sum_{i=1}^m p^{v, C}_i=1$.
Dukhovny showed that Marichal's entropy  can be represented as 
an average of Shannon entropies of all probabilities  $p^{v, C}$ such that
 $C\in \mathcal{C}(2^N)$: 

$$H_M(v)=\frac{1}{n!} \sum_{{C}\in \mathcal{C}(2^N)}\ H_S(p^{v, C}).$$ 
Remark that $|\mathcal{C}(2^N)|=n!$. 


\section{Lattices and related ordered structures }
In this section, we investigate the relations between lattices and set 
systems. In particular we introduce a general class of sets systems called regular set systems, and also consider known classes of set systems called convex 
geometries and antimatroids. 
\begin{defn}[lattice]
Let $(L, \leq)$ be a partially  ordered set, i.e. $\leq$ is a binary relation on $L$ being reflexive, antisymmetric and transitive. 
$(L, \leq)$ is called a lattice if for all $x, y \in L$, 
the least upper bound $x\vee y$ and 
the greatest lower bound $x\wedge y$ of $x$ and $y$ exist. 
\end{defn}

Let $L$ be a lattice. If $\bigvee S$ and $\bigwedge S$ exist for all $S\subseteq L$, then 
$L$ is called a {\it complete lattice}. 
$\bigvee L$ and $\bigwedge L$ are 
called the \emph{top element} and the \emph{bottom element} of $L$ and written 
$\top$ and $\bot$, respectively. 
We denote a complete lattice by $(L, \leq, \vee, \wedge, \bot, \top)$. 
If $L$ is a finite set, 
then $L$ is a complete lattice. 

The \emph{dual} of a statement about lattices phrased in terms of $\vee$ and 
$\wedge$ is obtained by interchanging  $\vee$ and 
$\wedge$. If a statement about lattice is true, then the dual statement is 
also true. This fact is called the \emph{duality principle}.

\begin{defn}[$\vee$-irreducible element]
An element $x\in (L, \leq)$ is 
$\vee$-{\it irreducible} if  for all $a, b \in L$, $x \ne \bot$ and $x =a \vee b$ implies 
$x=a$ or $x=b$.  
\end{defn}
The dual of a $\vee$-irreducible element is called a 
$\wedge$-{\it irreducible element}, which satisfies that if for all 
$a, b\in L$, $x\ne \top$ and $x=a \wedge b$ implies $x=a$ or $x=b$. 
We denote the set of all
$\vee$-irreducible elements of $L$ by $\mathcal{J}(L)$ and 
the set of all $\wedge$-irreducible elements 
of $L$ by  $\mathcal{M}(L)$.  


The mapping $\eta$ for any $a\in L$, defined by
$$\eta(a):=\{x \in \mathcal{J}(L)\ |\ x\leq a\} $$
is a lattice-isomorphism of $L$ onto $\eta(L):=\{\eta(a)\mid a\in L\}$, 
that is, $(L, \leq) \cong (\eta(L), \subseteq)$. 
Obviously $(\mathcal{J}(L), \eta(L))$ is a 
set system (see Section 6.1).  

We say $a$ is {\it covered} by $b$, and write $a\prec b$ or $b\succ a$, if 
$a<b$ and $a\leq x<b$ implies $x=a$. 

\begin{defn}[maximal chain of lattice]
$C=(c_0, c_1, \ldots, c_m)$ is a maximal 
chain of $(L, \leq)$ if $c_i \in L, i=0, \ldots, m$, and 
$\bot=c_0 \prec c_1 \prec \cdots \prec c_m =\top$.  
\end{defn}

We denote the set of all maximal chains of $L$ by 
$\mathcal{C}(L)$. 

We introduce the regular property for set systems. 

\begin{defn}[regular set system]
Let $(N, \mathfrak{S})$ be a set system. 
We say that $\mathfrak{S}$ is a regular set system if for any $C\in \mathcal{C}(\mathfrak{S})$, the length of $C$ is $n$, i.e. $|C|=n+1$. 
\end{defn}

\begin{defn}[$\vee$-minimal regular] 
If $(L, \leq)$ satisfies that 
the length of $C$
is $|\mathcal{J}(L)|$, i.e. $|C|=|\mathcal{J}(L)|+1$, 
for any $C\in \mathcal{C}(L)$
then we say that $(\mathcal{J}(L), \eta(L))$ 
is $\vee$-minimal regular. 
\end{defn}

\begin{lem}\label{rrn}
If $(L, \leq)$ is $\vee$-minimal regular then 
$(\mathcal{J}(L), \eta(L))$ is a 
regular set system. 
\end{lem}
\begin{pf}
Since $\eta(L)$ is isomorphic to $L$, for any 
$C\in \mathcal{C}(\mathcal{J}(L))$, $|C|=|\mathcal{J}(L)|+1$ holds. 
\end{pf}

\begin{lem}\label{tdsl}
If $L$ is $\vee$-minimal regular, then for every maximal chains $C=
(c_0, c_1, \ldots, c_n)$, where $n=|\mathcal{J}(L)|$, it holds that 
$\eta(c_i)=\eta(c_{i-1})
\cup \{j\}$ for some $j\in \mathcal{J}(L)$. 
\end{lem}

\begin{pf}
It suffices to show that $|\eta(c_i)|=i$. Suppose that there exists $i_0$ 
such that $|\eta(c_{i_0})|>i_0$. 
Since $|C|=n$ and $|\eta(c_i)\setminus \eta(c_{i-1})|
\geqq 1$ for any $i=1, \ldots n$, there will be not enough $\vee$-irreducible elements to complete the chain. 
\end{pf}

\begin{defn}[convex geometry and antimatroid] \label{cg} 
Let  $(N, \mathfrak{S})$ be a set system. 
$\mathfrak{S}$ is called
a convex geometry of $N$ if
\begin{description}
		\item[~~\rm (i)] $\emptyset, N \in \mathfrak{S}$,  
		\item[~~\rm (ii)] for any $A, B \in \mathfrak{S}$, 
		$A\cap B\in \mathfrak{S}$, 
		\item[~~\rm (iii)] for any $A\in \mathfrak{S}\setminus \{N\}$, 
there exists $i \in N\setminus A$ such that $A\cup \{i\}\in \mathfrak{S}$. 
\end{description}
Let $(N, \mathfrak{S})$ be a convex geometry. The dual system 
of $(N, \mathfrak{S})$ defined by 
$\mathfrak{A}=\{N\setminus A\mid A\in \mathfrak{S}\}$, 
$(N, \mathfrak{A})$ is called 
antimatroid. 
\end{defn}

Following result can be found in \cite{Monjardet:1990}\cite{Monjardet:2003}. We give a proof for the sake of completeness.

\begin{lem}\label{crr}
If $(N, \mathfrak{S})$ is a convex geometry or an antimatroid, then 
$\mathfrak{S}$ is a regular set system. 
\end{lem}

\begin{pf}Let $\mathfrak{S}$ be a convex geometry. 
Suppose that there exists $C=(c_0, c_1, \ldots, c_k)
\in \mathcal{C}(\mathfrak{S})$ such that 
$|C|< n+1$. Then we can take $c_i\in C$ which satisfies $|c_i\setminus 
c_{i-1}|>1$. 
We have $c_{i-1}\subsetneq c_i\subseteq N$,  and 
by (iii) of Definition  \ref{cg}, we can take $j_1, \ldots, 
j_t\in N$ such that 
$c_{i-1}\cup \{j_1\}, c_{i-1}\cup \{j_1, j_2\}, \ldots, c_{i-1}\cup 
\{j_1, \ldots j_{t}\}\in \mathfrak{S}$ and 
$c_{i-1}\cup \{j_1, \ldots j_{t}\}=N$, so that in these 
elements there exists 
an element $c$ such that $|c\cap c_i|=|c_i|-1$. By (ii), 
$c\cap c_i\in \mathfrak{S}$ and 
$c_{i-1}\subsetneq (c\cap c_i) \subsetneq c_i$, which 
contradicts the fact that $C$ is maximal. Hence $|C|\geqq n+1$.  
On the other hand, obviously, for any $C\in \mathcal{C}(\mathfrak{S})$, $|C| 
\leqq n+1$, 
hence  $|C|=n+1$.
And by the duality principle, the antimatroid is also a regular set system. 
\end{pf}

Convex geometries and antimatroids are complete lattices 
$(\mathfrak{S}, \subseteq, \vee, \cap, N, \emptyset)$ 
and $(\mathfrak{S}, \subseteq, \cup, \wedge, N, \emptyset)$, respectively, where $x\vee y :=\bigcap \{z\in \mathfrak{S} \mid x\cup y \subseteq z\}$ 
and $x\wedge y :=\bigcup \{z\in \mathfrak{S} \mid x\cap y \subseteq z\}$. 

\begin{figure}[htb]
\begin{center}
\unitlength 1cm
\begin{picture}(5,5)  
\put(0,1){\includegraphics[width=4cm,height=3.6cm,clip]{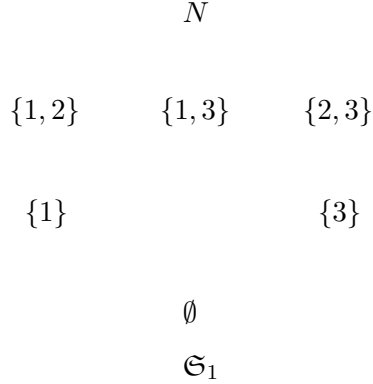}}
{\small
\put(2.0,4.7){$N$}
\put(-0.1,2.05){$\{1\}$}
\put(3.8,2.05){$\{3\}$}
\put(-0.3,3.4){$\{1,2\}$}
\put(1.7,3.4){$\{1,3\}$}
\put(3.6,3.4){$\{2,3\}$} 
\put(2.0,0.7){$\emptyset$}
\put(2.0,0){$\mathfrak{S}_1$}
}
\end{picture}
\end{center}
\caption{Antimatroid}
\label{am}
\end{figure}

\begin{lem} \label{car}
If $(N, \mathfrak{S})$ is a convex geometry, 
then $|\mathcal{J}(\mathfrak{S})|=n$.
Similarly, if $(N, \mathfrak{S})$ is an antimatroid, 
then $|\mathcal{M}(\mathfrak{S})|=n$. 
\end{lem}
\begin{pf}

Suppose that $\mathfrak{S}$ is a convex geometry. 
By Lemma \ref{crr}, for any $a \in \mathfrak{S}$, 
we have $a\setminus \underbar{\it a}\in N$, where $\underbar{\it a}\prec a$. 
And for any $b, c\in \mathcal{J}(\mathfrak{S})$ such that $b\neq c$, we have 
$b\setminus \underbar{\it b}\ne c\setminus \underbar{\it c}$, 
because when $\underbar{\it b}=\underbar{\it c}$,  we have 
$b\setminus \underbar{\it b}\ne c\setminus \underbar{\it c}$ 
obviously, and when 
$\underbar{\it b}\ne\underbar{\it c}$, 
$b\setminus \underbar{\it b}= c\setminus \underbar{\it c}$  means 
$b\cap c \supseteq b\setminus \underbar{\it b}$ and $b=(b\cap c )
\cup \underbar{\it b}$, which contradicts that 
$b$ is a $\vee$-irreducible element, 
so that $|\mathcal{J}(\mathfrak{S})|\leqq n$. On the other hand, for any chain $C=(c_0, \ldots, c_n)\in \mathcal{C}(\mathfrak{S})$, $|\eta(c_i)|>|\eta(c_{i-1})|$ so that 
$n\leqq |\eta(\top)|=|\mathcal{J}(\mathfrak{S})|$. Therefore $|\mathcal{J}(\mathfrak{S})|=n$. By the duality principle, the same is true for 
antimatroids. 
\end{pf}
For example, $\mathfrak{S}_1$  in Fig. \ref{am} is an antimatroid 
and  a regular set system of $N=\{1, 2, 3\}$.  
$|\mathcal{J}(\mathfrak{S}_1)|=|\{1, 3, 12, 23\}|=4$ and 
$|\mathcal{M}(\mathfrak{S}_1)|=|\{12, 13, 23\}|=3$. 

\begin{figure}[htb]
\begin{center}
\unitlength 1cm
\begin{picture}(5,5)  
\put(0,1){\includegraphics[width=4cm,height=3.6cm,clip]{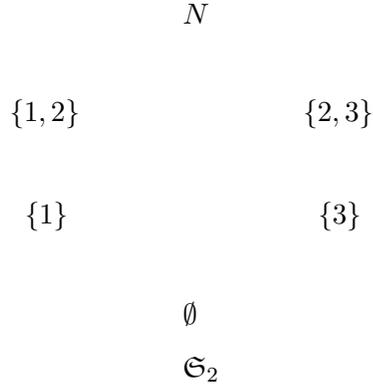}}
{\small
\put(2.0,4.7){$N$}
\put(-0.1,2.05){$\{1\}$}
\put(3.8,2.05){$\{3\}$}
\put(-0.3,3.4){$\{1,2\}$}
\put(3.6,3.4){$\{2,3\}$} 
\put(2.0,0.7){$\emptyset$}
\put(2.0,0){$\mathfrak{S}_2$}
}
\end{picture}
\end{center}
\caption{Regular set system}
\label{rss}
\end{figure}

If $\mathfrak{S}$ is a regular set system, it does not necessarily 
hold that $|\mathcal{J}(\mathfrak{S})|=n$ nor 
$|\mathcal{M}(\mathfrak{S})|=n$. 
Consider the lattice 
$\mathfrak{S}_2$ in Fig.~\ref{rss}. 
$\mathfrak{S}_2$ is a regular set system of $\{1, 2, 3\}$, but 
$\mathcal{J}(\mathfrak{S})=\mathcal{M}(\mathfrak{S})=\{1, 3, 12, 23\}$. 

{\bf Remark}\ 
A segment $[a, b]$ of $L$, for $a, b \in L$, is the set if all elements $x$ 
which satisfy $a\leq x\leq b$. If $\mathfrak{S}$ is a regular set system 
then $\mathfrak{S}$ satisfies \emph{the Jordan-Dedekind chain condition}, 
that is, all maximal chains in any segments of $\mathfrak{S}$ have the same length. 
The converse does not hold. For instance, $(N, \mathfrak{S})=(\{1,2,3\},\{\emptyset, \{1,2\}, \{3\}, N\})$ satisfies the Jordan-Dedekind chain condition but is not 
a regular set system. Incidentally, $\mathfrak{S}$ is $\vee$-minimal regular. 
Similarly, If $L$ is a convex geometry or an antimatroid, then $L$ satisfies 
the Jordan-Dedekind chain condition, but the converse does not hold. 
For instance, consider the lattice $(\mathfrak{S}, \subseteq)=(\{1,2,3,4\},\{\emptyset, \{1\}, \{1,2\}, \{3\},
 \{3,4\},  N\})$
\section{Capacity on lattice}

\begin{defn}[capacity on lattice]
A mapping $v : L\to [0,1]$ is a capacity on $L$ if it satisfies 
$v(\bot)=0, v(\top)=1$ and for any $x, y \in L$, 
$v(x)\leqq v(y)$ whenever $x\leq y$.  
\end{defn}

\begin{defn}[cardinality-based capacity]
A capacity on $(N, \mathfrak{S})$ is \\
cardinality-based 
 if $v(A)$ depends 
only on $|A|$ for any $A\in \mathfrak{S}$. 
\end{defn}

\begin{defn}[additive uniform capacity]
The additive uniform capacity on $(N, \mathfrak{S})$ is defined by 
$$\ds v^\ast (A):=\frac{|A|}{n}$$ 
for any $A \in \mathfrak{S}$. 
\end{defn}

Uniform capacities and the additive uniform capacity 
can be defined on any lattice $L$  by  putting 
$|x|:=|\eta(x)|$ for any $x\in L$. 

Faigle and Kern generalized the Shapley value to that of a game on a 
lattice \cite{FK}, and 
Bilbao defined it 
for  games on convex geometries \cite{BE} 
and on antimatroids \cite{ABBJ}.   

\begin{defn}[Bilbao and Edelman's Shapley value]
Let $v$ be a game on a convex geometry
or an antimatroid $(N, \mathfrak{S})$. 
For $i\in N$, the Shapley value of $v$ is defined by
\begin{equation}\label{svoca}
\hspace{2cm}\phi_{i}(v):=\frac{1}{|\mathcal{C}(\mathfrak{S})|}\sum_{C\in \mathcal{C}(\mathfrak{S})\atop A, A\cup \{i\}\in C}(v(A\cup \{i\})-v(A)). 
\end{equation}
\end{defn}
When $\mathfrak{S}$ is a regular set sytem of $N$,  
we can also define the Shapley value of games 
on $\mathfrak{S}$ by 
(\ref{svoca}). 

By Lemma \ref{rrn}, regarding the lattice as a set system of 
$\mathcal{J}(L)$, we can also calculate the Shapley 
value of capacities on the regular lattice as follows.  

\begin{defn}[Shapley value on $L$ (cf. \cite{FK})] 
Suppose that $(L, \leq)$ is $\vee$-minimal regular  
and let 
$v$  be a capacity on $L$. 
For $x \in \mathcal{J}(L)$, the Shapley 
value of $v$ on $L$ is defined by
\begin{eqnarray}
\phi_x(v)&:=&\frac{1}{|\mathcal{C}(L)|} \sum_{C\in \mathcal{C}(L) \atop 
\eta(c_i)\setminus \eta(c_{i-1})=x} 
\ (v(c_i)-v(c_{i-1}))\nonumber \\
&=&\frac{1}{|\mathcal{C}(L)|} \sum_{C\in \mathcal{C}(L) \atop 
\eta(c_i)\setminus \eta(c_{i-1})=x} 
\ p^{v,C}_i, \label{svol}
\end{eqnarray}
where $C=(c_0, c_1, \ldots, c_n)$ and $n=|\mathcal{J}(L)|$. 
\end{defn}

By Lemma \ref{tdsl}, if $L$ is $\vee$-minimal regular, 
for any $C\in \mathcal{C}(L)$, 
$\eta(a)\setminus \eta(b)\in \mathcal{J}(L)$ 
for any $a, b \in C$ such that $a \prec b$. Hence formulas 
(\ref{svol}) are well-defined. 
Similarly, if 
$L$ satisfies the following property: 
\begin{description}
	\item[($\wedge$-minimal regular)] 
any $C\in \mathcal{C}(L)$, the length of $C$ is $|\mathcal{M}(L)|$, i.e. $|C|=|\mathcal{M}(L)|+1$, 
\end{description}
 for $x\in \mathcal{M}(L)$,  we can calculate the Shapley value of capacities on $L$ in a similar manner as follows. 
For $x\in \mathcal{M}(L)$, the Shapley value of $v$ on $L$ is defined by 
\begin{eqnarray*}
\phi_x(v)&:=&\frac{1}{|\mathcal{C}(L)|} \sum_{C\in \mathcal{C}(L) \atop 
\eta^d(c_{i-1})\setminus \eta^d(c_{i})=x} 
\ (v(c_i)-v(c_{i-1}))\nonumber \\
&=&\frac{1}{|\mathcal{C}(L)|} \sum_{C\in \mathcal{C}(L) \atop 
\eta^d(c_{i-1})\setminus \eta^d(c_{i})=x} 
\ p^{v,C}_i, 
\end{eqnarray*}
where $C=(c_0, c_1, \ldots, c_n)$, $n=|\mathcal{M}(L)|$ and $\eta^d(a):=\{x\in \mathcal{M}(L)\mid x\geq a\}$. 
We have $(L, \leq)\cong(\mathcal{M}(L), \eta^d(L))$. 

If $L$ is both $\vee$ and $\wedge$-minimal regular, 
then we can use both 
$\mathcal{J}(L)$ and 
$\mathcal{M}(L)$ for calculating the Shapley value. 
However  $\mathcal{J}(L)$ is better, 
because  elements of $\mathcal{J}(L)$ are in general easier 
to interpret (cf. Section \ref{mcg}).

\section{Entropy of capacities on lattices and set systems}

In this 
section, we suppose that $(N, \mathfrak{S})$ is a regular set system 
and let $v$ and $u$ be capacities on $(N, \mathfrak{S})$.  

\begin{defn}[entropy] 
Let $v$ be a capacity on $\mathfrak{S}$. 
The entropy of $v$ 
is defined by
\begin{eqnarray}
\hspace{3cm}H(v)
&:=&\frac{1}{|\mathcal{C}(\mathfrak{S})|}\sum_{C\in \mathcal{C}(\mathfrak{S})} 
H_S(p^{v,C}), \label{erl}
\end{eqnarray}
where $C=(c_0, c_1, \ldots, c_n)$. 
\end{defn}

\begin{defn}[relative entropy] 
Let $v$ and  $u$ be capacities on $\mathfrak{S}$. 
The relative entropy of $v$ to 
$u$ is defined by
\begin{eqnarray}
\hspace{3cm}H(v; u)
&:=&\frac{1}{|\mathcal{C}(\mathfrak{S})|}\sum_{C\in \mathcal{C}(\mathfrak{S})} H_S(p^{v, C}; p^{u, C}). \label{rrr}
\end{eqnarray}
\end{defn}
Let $v$ and $u$ be  capacities on $L$. If $L$
is $\vee$ or $\wedge$-minimal regular, then regarding $L$ as a set system 
$(\mathcal{J}(L), \eta(L))$ or $(\mathcal{M}(L), \eta^d(L))$, 
we can also define the entropy $H(v)$  and the relative entropy $H(v; u)$ 
as follows.  
\begin{eqnarray}
\hspace{3cm}H(v)
&:=&\frac{1}{|\mathcal{C}(L)|}\sum_{C\in \mathcal{C}(L)} 
H_S(p^{v,C}), \label{eol}\\
H(v; u)
&:=&\frac{1}{|\mathcal{C}(L)|}\sum_{C\in \mathcal{C}(L)} H_S(p^{v, C}; p^{u, C}). \label{reol}
\end{eqnarray}

We can consider that $H(v)$ is an average of Shannon entropies, 
and also that $H(u; v)$ is an average of Shannon relative entropies. Therefore they satisfy several properties which are required for entropies 
(cf. \cite{KM}). 

\begin{prop}For any $v$, $H(v)$ is a continuous function, and 
$0\leqq H(v)\leqq \log n$, with equality on left side if and only if $v$ is $\{0, 1\}$-valued capacity, and 
with equality on right side if and only if $v$ is the additive 
uniform capacity $v^\ast$. 
\end{prop}
\begin{pf}
The continuity is obvious. For any probability $p$, 
$H_S(p)\geqq 0$, so that $H(v)\geqq 0$ holds. 
$H_S(p)= 0$ if and only if $p$ is deterministic, i.e. there exists $i$ 
such that $p_i=1$ and otherwise $p_j=0$. Hence 
for all $C\in\mathcal{C}(\mathfrak{S})$, $p^{v, C}_i$ takes value only $0$ or $1$, 
which means that for all $A\in \mathfrak{S}$, $v(A)$ takes value only $0$ or $1$.  
Similarly, $H_S(p)\leqq \log{n}$, so that an average of $H_S(p)$ is 
dominated by $\log{n}$. $H_S(p)= \log n$ if and only if for all $i$, 
$p_i=1/n$, hence for all $a\in \mathfrak{S}$, 
$v(A)=|A|/n$, 
which completes the proof. 
\end{pf}

\begin{prop}
For any uniform capacity $v$ on $\mathfrak{S}$, we have 
$$H(v)=H_S(p^{v, C})$$
for any $C\in \mathcal{C}(\mathfrak{S})$. 
\end{prop}
\begin{pf}
In this case, for all $C\in \mathcal{C}(\mathfrak{S})$, $p^{v, C}$ is the same 
probability distribution, 
hence we have 
\begin{eqnarray*}
H(v)&=&\frac{1}{|\mathcal{C}(\mathfrak{S})|} \sum_{C\in \mathcal{C}(\mathfrak{S})} H_S(p^{v, C})\\
&=&\frac{1}{|\mathcal{C}(\mathfrak{S})|} \ |\mathcal{C}(\mathfrak{S})| \  H_S(p^{v, C})=H_S(p^{v, C}).
\end{eqnarray*}
\end{pf}

Define $v_{\lambda}:=(1-\lambda)v+\lambda v^\ast$
for $0<\lambda< 1$. Then for any $v (\not\equiv v^\ast)$,  
$H(v_\lambda)$ is strictly increasing toward 
 the additive uniform capacity $v^\ast$. 
\begin{prop}\quad  
For any $v(\not\equiv v^\ast)$, 
$H(v_\lambda)$ is a strictly 
increasing function of $\lambda$. 
\end{prop}
\begin{pf}
We show that $\ds \frac{ dH_S ( p^{v_{\lambda}, C}) }{d\lambda}>0$
for any ${C}\in \mathcal{C}(\mathfrak{S})$ 
such that $p^{v, C}\not\equiv p^\ast$. 
\begin{eqnarray*}
H_S(p^{v_\lambda, C})&=&\sum_{i=1}^n h[v_{\lambda}(c_i)-v_\lambda(c_{i-1})]\\
&=&\sum_{i=1}^n h\left[p^{v, C}_i+\lambda\left(\frac{1}{n}-p^{v, C}_i \right)\right],
\end{eqnarray*}
where $C=(c_0, \ldots, c_n)$ and $p^{v, C}_i:=v(c_i)-v(c_{i-1})$.
\begin{eqnarray*}
\frac{dH_S(p^{v_\lambda, C}_i)}{d\lambda}
&=&\sum_{i=1}^n \left(p^{v, C}_i-\frac{1}{n}\right)
\left(1+\log\left(p^{v, C}_i+\lambda\left(\frac{1}{n}-p^{v, C}_i
\right)\right)\right)\\
&=&\sum_{i=1}^n \left(p^{v, C}_i-\frac{1}{n}\right)
\left(\log\left(p^{v, C}_i+\lambda\left(\frac{1}{n}-p^{v, C}_i\right)\right)\right)
\end{eqnarray*}
If $1/n\geqq p^{v, C}_i$, then 
$$p^{v, C}_i+\lambda\left(\frac{1}{n}-p^{v, C}_i\right)\in
\left[p^{v, C}_i, \frac{1}{n}\right)$$
and otherwise, that is, $1/n< p^{v, C}_i$, we have 
$$p^{v, C}_i+\lambda\left(\frac{1}{n}-p^{v, C}_i\right)\in
\left( \frac{1}{n}, p^{v, C}_i\right),$$
so that we have 
\begin{eqnarray*}
\frac{dH_S(p^{v_\lambda, C}_i)}{d\lambda}
&>&\sum_{i:1/n\geqq p^{v, C}_i}\left(\frac{1}{n}-p^{v, C}_i\right)
\log\frac{1}{n}+\sum_{i:1/n< p^{v, C}_i}
\left(\frac{1}{n}-p^{v, C}_i\right)
\log\frac{1}{n}\\
&=&\log\frac{1}{n}\sum_{i=1}^n\left(\frac{1}{n}-p^{v, C}_i\right)
=\log\frac{1}{n}\left(1-\sum_{i=1}^n p^{v, C}_i\right)
=0
\end{eqnarray*}
Since $v\not\equiv v^\ast$, there exist at least an $C \in \mathcal{C}(\mathfrak{S})$  
such that $p^{v, C}\not\equiv p^{v^\ast, C}$, therefore 
$$H(v)=\frac{1}{|\mathcal{C}(\mathfrak{S})|}\sum_{C\in \mathcal{C}(\mathfrak{S})} H_S(p^{v_\lambda, C}_i)$$ is a strictly increasing function of $\lambda$. 
\end{pf}
\begin{prop}\quad \label{re}
$H(v; u)\geqq 0$ and that  $H(v; u)=0$ if and only if 
$v\equiv u$.
\end{prop}
\begin{pf}
Non-negativity is obvious by $H_S(p; q) \geqq 0$. And 
$H(v; u)=0$ if and only if 
$H_S(p^{v, C}; p^{u, C})=0$ for all $C\in \mathcal{C}(\mathfrak{S})$, 
which is true if and only if 
$p^{v, C} \equiv p^{u, C}$ for all $C\in \mathcal{C}(\mathfrak{S})$,  
which means $v\equiv u$. 
\end{pf}
\begin{prop}\quad 
Let $v\not\equiv u$ and $v^u_\lambda :=\lambda u +(1-\lambda) v$. Then $H(v^u_\lambda ; u)$ is 
a strictly decreasing function of $\lambda$. 
\end{prop}
\begin{pf}\quad 
We show that 
$$\frac{d H_s(p^{v^u_\lambda, C}; p^{u, C})}{d\lambda}<0$$
for any $C \in \mathcal{C}(\mathfrak{S})$ such that $p^{v^u_\lambda, C}\not\equiv p^{u, C}$. 
\begin{eqnarray*}
H(p^{v^u_\lambda, C}; p^{u, C})
&=&\sum_{i=1}^n p^{v^u_\lambda, C}_i \log \frac{p^{v^u_\lambda, C}_i}{p^{u, C}_i}\\
&=&\sum_{i=1}^n h\left[\lambda(p^{u, C}_i-p^{v, C}_i) + p^{v, C}_i; p^{u, C}_i\right]
\end{eqnarray*}
\begin{eqnarray*}
\frac{d H(p^{v^u_\lambda, C}; p^{u, C})}{d \lambda}&=&
(p^{u, C}_i-p^{v, C}_i) \left(\log\frac{\lambda(p^{u, C}_i-p^{v, C}_i)+p^{v, C}_i}{p^{u, C}_i} +1\right).
\end{eqnarray*}
If $p^{u, C}_i \geqq p^{v,C}_i$, then 
$$\frac{\lambda(p^{v, C}_i-p^{v, C}_i)+p^{v, C}_i}{p^{u, C}_i}\in 
\left[\frac{p^{v, C}_i}{p^{v, C}_i}, 1\right), $$
and otherwise, that is, $p^{u, C}_i < p^{v, C}_i$
$$\frac{\lambda(p^{v, C}_i-p^{v, C}_i)+p^{v, C}_i}{p^{u, C}_i}\in 
\left(1, \frac{p^{v, C}_i}{p^{v, C}_i}\right), $$
so that we have 
\begin{eqnarray*}
\frac{d H(p^{v^u_\lambda, C}; p^{u, C})}{d \lambda}&<&
\sum_{i : p^{u, C}_i \geqq p^{v, C}_i}(p^{u, C}_i - p^{v, C}_i)
+\sum_{i : p^{u, C}_i < p^{v, C}_i}(p^{u, C}_i - p^{v, C}_i)\\
&=&\sum_{i=1}^n p^{u, C}_i -\sum_{i=1}^n p^{v, C}_i)=0
\end{eqnarray*}
Since $v\not\equiv u$, there exists at least un $C\in \mathcal{C}(\mathfrak{S})$ such that 
$p^{v, C} \not\equiv p^{u, C}$, therefore 
$$H(v; u)=\frac{1}{|\mathcal{C}(\mathfrak{S})|}\sum_{C\in \mathcal{C}(L)} H_S(p^{v, C}; 
p^{u, C})$$
is a strictly decreasing function of $\lambda$. 
\end{pf}

\section{Examples}
In this section, we show several examples. 
Most games and capacities which appear in applications are 
particular capacities on regular set systems.  

\subsection{Regular lattice}
$L_1$ in Fig. \ref{gcg} is $\vee$-minimal regular, and is also isomorphic to  
a convex geometry. 
\begin{figure}[htb]
\begin{center}
\unitlength 1cm
\begin{picture}(10.5,5)  
\put(0,1){\includegraphics[width=4cm,height=3.6cm,clip]{l_cg.bmp}}
\put(2.0,4.7){$a$}
\put(0,2.05){$d$}
\put(2.0,2.05){$e$}
\put(4.0,2.05){$f$}
\put(0,3.4){$b$}
\put(4.0,3.4){$c$} 
\put(2.0,0.7){$g$}
\put(2.0,0){$L_1$}
\put(6,1){\includegraphics[width=4cm,height=3.6cm,clip]{l_cg.bmp}}
\put(7.4,4.7){$\{d,e,f\}$} 
\put(5.9,2.05){$\{d\}$}
\put(7.8,2.05){$\{e\}$}
\put(9.8,2.05){$\{f\}$}
\put(5.7,3.4){$\{d,e\}$}
\put(9.6,3.4){$\{e,f\}$} 
\put(8.0,0.7){$\emptyset$}
\put(8.0,0){$\eta(L_1)$}

\end{picture}
\end{center}
\caption{}
\label{gcg}
\end{figure}

In fact, $\mathcal{J}(L_1)=\{ d, e, f\}$, and 
$L_1$ is also represented by 
$\eta(L_1)$. 
$\mathcal{C}(\eta(L_1))=\{(\emptyset, \rm d, de, def), 
(\emptyset, \rm e, de, def), (\emptyset, \rm e, ef, def), (\emptyset, \rm f, ef, def)\}$.  
Let $v$ be a capacity on  $L_1$. 
Then the Shapley values and 
the entropy of $v$ on $L_1$ are as follows. 
\begin{eqnarray*}
\phi_{d}(v)&=&\frac{1}{4}(v(d)-v(g))+\frac{1}{4}(v(b)-v(e))+\frac{1}{2}(v(a)-v(c))\\
\phi_{e}(v)&=&\frac{1}{2}(v(e)-v(g))+\frac{1}{4}(v(b)-v(d))+\frac{1}{4}(v(c)-v(f))\\
\phi_{f}(v)&=&\frac{1}{4}(v(f)-v(g))+\frac{1}{4}(v(c)-v(e))+\frac{1}{2}(v(a)-v(b))
\end{eqnarray*}
and 
\begin{eqnarray*}
H(v)&=&\frac{1}{4}h[v(d)-v(g)]+\frac{1}{4}h[v(b)-v(e)]+\frac{1}{2}h[v(a)-v(c)]\\&&+\frac{1}{2}h[v(e)-v(g)]+\frac{1}{4}h[v(b)-v(d)]+\frac{1}{4}h[v(c)-v(f)]\\
&&+\frac{1}{4}h[v(f)-v(g)]+\frac{1}{4}h[v(c)-v(e)]+\frac{1}{2}h[v(a)-v(b)]. 
\end{eqnarray*}

\subsection{Distributive lattice}
$(L, \leq)$ is said to be distributive if it satisfies the distributive law, 
$a \wedge (b\vee c) =(a\wedge b) \vee (a \wedge c)$ for any $a, b, c \in L$. 
If $(L, \leq)$ is distributive then 
$(L, \leq)$ is also $\vee$ and $\wedge$-minimal regular. 
Remark that a regular set system, even the convex geometry and  
the  antimatroid are not necessarily distributive (cf. Fig. 2, Fig. 3). 


\subsection{Capacity on $2^N$ (classical capacity)}
The classical capacity is a monotone function on the Boolean lattice $2^N$. 
$2^N$ is a distributive lattice and 
also a complemented lattice, i. e. for any $A\in 2^N$, there exists 
a complement $B\in 2^N$ such that $A\wedge B =\bot=\emptyset$ 
and $A\vee B =\top=N$.
For any capacity on $2^N$, (\ref{svoca}) is equals to the Shapley value (\ref{sv}), 
and our entropies (\ref{erl}) and (\ref{rrr}) are equal to Marichal's entropy 
(\ref{me}) (cf. Section 2). 
\subsection{Bi-capacity {\rm \cite{bi1}\cite{bi2}}}\label{bi}
A bi-capacity is a monotone function on 
$\mathcal{Q}(N):=\{(A, B)\in 2^N \times 2^N \mid A\cap B=\emptyset\}$ 
which satisfies that $v(\emptyset, N)=-1, v(\emptyset, \emptyset)=0$ and 
$v(N, \emptyset)=1$. 
For any $(A_1, A_2), (B_1, B_2)\in \mathcal{Q}(N)$, $(A_1, A_2)\sqsubseteq 
(B_1, B_2)$ iff $A_1\subseteq B_1$ and $A_2\supseteq B_2$. 
$\mathcal{Q}(N)\cong 3^N$. 
It can be shown that $(\mathcal{Q}(N), \sqsubseteq )$ is a finite distributive lattice. Sup and inf 
are given by 
$(A_1, A_2)\vee (B_1, B_2)=(A_1\cup B_1, A_2\cap B_2)$ and 
$(A_1, A_2)\wedge (B_1, B_2)=(A_1\cap B_1, A_2\cup B_2)$, 
and we have 
$$\mathcal{J}(\mathcal{Q}(N))=\{(\emptyset, N\setminus\{i\}), i\in N\} \cup 
\{(\{i\}, N\setminus\{i\}), i\in N\},$$
where $i\in N$. 
Normalizing $v$ by $v'  :\mathcal{Q}(N) \to [0, 1]$ such that 
$$v':=\frac{1}{2}v +\frac{1}{2},$$ 
we can regard $v$ as a capacity on $\mathcal{Q}(N)$. Then, applying 
(\ref{svol}) 
and  (\ref{eol}), we have  
\begin{eqnarray*}
\phi^+_i(v')&:=&\phi_{(\{i\}, N\setminus\{i\})}(v')\\
&=&\sum_{A\subseteq N\setminus \{i\}\atop B\subseteq N\setminus (A\cup \{i\})}
\gamma^n_{|A|, |B|}
\left( v'(A\cup \{i\}, B)-v'(A, B)\right),\\
\phi^-_{i}(v')&:=&
\phi_{(\emptyset, N\setminus\{i\})}(v')\\
&=&\sum_{A\subseteq N\setminus \{i\}\atop B\subseteq N\setminus (A\cup \{i\})}
\gamma^n_{|A|, |B|}
\left(v'(B, A)- v'(B, A\cup\{i\})\right)
\end{eqnarray*}
and
\begin{eqnarray*}
\lefteqn{H(v')=\sum_{i=1}^n \sum_{A\subset N\setminus x_i \atop B\subset N\setminus (A\cup \{i\})}\gamma^n_{|A|, |B|} \left(h\left[ v'(A\cup \{i\}, B)-v'(A,B)\right]\right.}&&\\
&&\left. \hspace{6cm}+h
\left[v'(B, A)-v'(B, A\cup \{i\})\right]\right).
\end{eqnarray*}
where $\ds \gamma^n_{k, \ell}:=\frac{(n-k+\ell-1)!\ (n+k-\ell)!\ 2^{n-k-\ell}}{(2n)!}$, and $h(x):=-x \log x$. 

$\phi^+_{i}$ and $\phi^-_{i}$ mean positive and negative degrees of $i$'s contribution to $v$, respectively, hence the contribution of $i$ to $v$ is given by 
$\phi_{i}(v):=\phi^+_{i}(v)+\phi^-_{i}(v)$. 
$\gamma^n_{|A|,|B|}$ is the rate of the number of chains which contain 
$(A\cup\{i\}, B)$ and $(A,B)$. In fact, 
\begin{eqnarray*}
\lefteqn{\left|\{C\in \mathcal{C}(\mathcal{Q}(N))\mid C\ni (A\cup\{i\}, B), (A,B)\}\right|}\hspace{4cm}\\
&=&\frac{(n+|A|-|B|)!}{(2!)^{|A|}}\cdot \frac{(n-|A|+|B|-1)!}{(2!)^{|B|}}
\end{eqnarray*}
and $|\mathcal{C}(\mathcal{Q}(N))|=(2n)!/(2!)^n$. 
These Shapley values are different from those in \cite{bi1}. 

\subsection{Multichoice game}\label{mcg}
Multichoice games have been proposed by Hsiao and Raghavan \cite{hsra93}. They 
have been proposed also independently in the context of capacities by Grabisch and Labreuche \cite{GL}, under the name k-ary capacities. 

\begin{figure}[htb]
\begin{center}
\unitlength 1cm
\begin{picture}(5,4)  
\put(0.1,0){\includegraphics[width=4.5cm,height=3.5cm,clip]{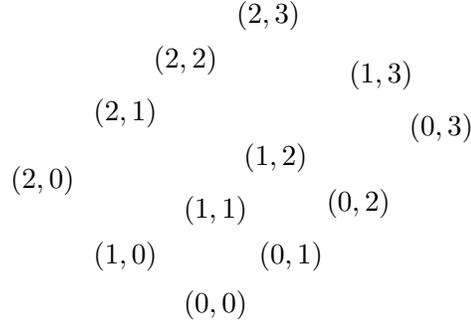}}
\put(1.6,-0.25){\small $(0,0)$}
\put(0.4,0.4){\small $(1,0)$}
\put(2.6,0.4){\small $(0,1)$}
\put(1.6,1){\small $(1,1)$}
\put(-0.7,1.4){\small $(2,0)$}
\put(3.5,1.1){\small $(0,2)$}
\put(0.4,2.3){\small $(2,1)$}
\put(2.4,1.7){\small $(1,2)$}
\put(1.2,3.0){\small $(2,2)$}
\put(4.6,2.1){\small $(0,3)$}
\put(3.8,2.8){\small $(1,3)$}
\put(2.3,3.6){\small $(2,3)$}
\end{picture}
\end{center}
\caption{$2$-players game}
\label{mcgf}
\end{figure}

Let $N:=\{0, 1,  \ldots. n\}$ be a set of players, and let $L:=
L_1\times\cdots \times L_n$, where 
($L_i, \leq_i)$ is a totally ordered set $L_i=
\{0, 1, \ldots, \ell_i\}$ such that $0\leq_i 1\leq_i \cdots \leq_i \ell_i$. 
Each $L_i$ is the set of choices of player $i$. 
$(L, \leq)$ is a regular lattice.  
For any $(a_1, a_2, \ldots, a_n), 
(b_1, b_2, \ldots, b_n) \in L$, 
$(a_1, a_2, \ldots, a_n) \leq 
(b_1, b_2, \ldots, b_n)$ iff $a_i \leq_i b_i$ for all $i=1, \ldots, n$. 
We have
$$\mathcal{J}(L)=\{(0, \ldots, 0, a_i, 0, \ldots, 0) \mid a_i\in 
\mathcal{J}(L_i)=L_i\setminus \{0\}\}$$
and $|\mathcal{J}(L)|=\sum_{i=1}^n \ell_i$. 
The lattice 
in Fig. \ref{mcgf} is 
an example of a product lattice, which 
represents a $2$-players game. 
Players $1$ and $2$ can choose among $3$ and $4$ choices. 
Let $v$ be a capacity on $L$, that is,  
$v(0, \ldots, 0)=0 $, 
$v(\ell_1, \ldots, \ell_n)=1$ and , for any $a, b\in
L$, $v(a)\leqq v(b)$ whenever $a\leq b$. 
In this case, applying (\ref{svol}) 
and  (\ref{eol}), we have  
\begin{eqnarray*}
\phi^j_i(v)&=&\phi_{(0, \ldots, 0, a_i=j>0, 0, \ldots, 0)}(v)\\
&=&\sum_{a\in L{/L_i}}\xi_i^{({\boldmath a}, j)}\left(v(\boldmath{a}, j)
-v(\boldmath{a}, j-1)\right)
\end{eqnarray*}
and 
\begin{eqnarray*}
H(v)&=&\sum_{i\in N\atop j\in L_i}\sum_{a\in L{/L_i}}\xi_i^{(\boldmath{a}, j)}\ h\left[v(\boldmath{a}, j)-v(\boldmath{a}, j-1)\right]
\end{eqnarray*}
where 
$L{/L_i}:=L_1\times \cdots \times L_{i-1} \times L_{i+1}\times \cdots \times L_n$, $(a, a_i):=(a_1, \ldots, a_{i-1}, a_i, a_{i+1}, \ldots, a_n)\in L$ such that 
$a\in L{/L_i}$ and $a_i \in L_i$, and 
$$\xi_i^{(a, a_i)}:=\left(\prod_{k=1}^n \left(\ell_k \atop a_k\right)\right) \cdot 
\left(\sum_{k=1}^n \ell_k \atop \sum_{k=1}^n a_k\right)^{-1}\cdot \frac{a_i}{\sum_{k=1}^n a_k}$$
and $h(x):=-x\log x$. 

$\phi^j_i(v)$ represents the contribution of player $i$ playing at level $j$ 
compared to level $j-1$,  where $j, j-1 \in \mathcal{J}(L_i)=L_i\setminus \{0\}$, hence player $i$'s overall contribution  is given by 
$$\phi_i(v)=\sum_{j=1}^{\ell_i} \phi^j_i(v). $$ 
$\xi^{(a, a_i)}_i$ is the rate of the number of chains which contain $(a, a_i)$
 and $(a, a_i-1)$. 
In fact, 
$$|\{C\in \mathcal{C}(L)\mid C\ni (a, a_i), (a, a_i-1)\}|=
\frac{\left(\sum_{k=1}^n a_k-1\right)!}{(\prod_{k=1}^n(a_k !))(a_i-1)!/(a_i!)}\cdot 
\frac{\left(\sum_{k=1}^n(\ell_k-a_k)\right)!}{\prod_{k=1}^n((\ell_k-a_k)!)}$$
and  $|\mathcal{C}(L)|=(\sum_{k=1}^n\ell_k)!/\prod_{k=1}^n(\ell_k!)$. 

Regarding a bi-capacity in Section \ref{bi} as a special case of 
multichoice game such that $n$ 
players and $\ell_i=2$ for all $i$ which is fixed a value $v'(\emptyset, \emptyset)=1/2$,  
we obtain the same 
Shapley values and the entropy. 

\section{Conclusion}
We have proposed a general definition of entropy for capacities defined on a large class of ordered structures we call regular set systems, which encompasses the original definition of Marichal for classical capacities. Regular set systems contain as particular important classes, distributive lattices, convex geometries and antimatroids. Hence our approach permits to define the entropy of multichoice games, also called $k$-ary capacities.

\end{document}